\documentclass[10pt,3p,twocolumn]{elsarticle}
\usepackage{amssymb}
\usepackage{url}
\usepackage{chngcntr}
\usepackage{booktabs}
\usepackage{caption}
\biboptions{super}

\begin{document}

\begin{frontmatter}

\title{\LARGE The emergent integrated network structure of scientific research}

\author[1]{Jordan D. Dworkin}
\author[1]{Russell T. Shinohara}
\author[2,3,4,5]{Danielle S. Bassett\corref{cor1}}

\address[1]{Department of Biostatistics, Epidemiology, and Informatics, Perelman School of Medicine, University of Pennsylvania, Philadelphia, PA, USA}
\address[2]{Department of Bioengineering, University of Pennsylvania, Philadelphia, PA, USA}
\address[3]{Department of Physics \& Astronomy, University of Pennsylvania, Philadelphia, PA, USA}
\address[4]{Department of Electrical \& Systems Engineering, University of Pennsylvania, Philadelphia, PA, USA}
\address[5]{Department of Neurology, University of Pennsylvania, Philadelphia, PA, USA}
\cortext[cor1]{Corresponding author:}
\ead{dsb@seas.upenn.edu}

\begin{abstract}
The practice of scientific research is often thought of as individuals and small teams striving for disciplinary advances. Yet as a whole, this endeavor more closely resembles a complex system of natural computation, in which information is obtained, generated, and disseminated more effectively than would be possible by individuals acting in isolation. Currently, the structure of this integrated and innovative landscape of scientific ideas is not well understood. Here we use tools from network science to map the landscape of interconnected research topics covered in the multidisciplinary journal \textit{PNAS} since 2000. We construct networks in which nodes represent topics of study and edges give the degree to which topics occur in the same papers. The network displays small-world architecture, with dense connectivity within scientific clusters and sparse connectivity between clusters. Notably, clusters tend not to align with assigned article classifications, but instead contain topics from various disciplines. Using a temporal graph, we find that small-worldness has increased over time, suggesting growing efficiency and integration of ideas. Finally, we define a novel measure of interdisciplinarity, which is positively associated with \textit{PNAS}'s impact factor. Broadly, this work suggests that complex and dynamic patterns of knowledge emerge from scientific research, and that structures reflecting intellectual integration may be beneficial for obtaining scientific insight.
\end{abstract}


\end{frontmatter}

\section{Introduction}
\label{intro}
The practice of scientific research represents the collective effort of humans to acquire information, generate insight, and disseminate knowledge. Although scientific inquiry has been carried out for centuries, the recent expansion of meta-data collection has allowed a robust body of literature to develop around the scientific study of science itself. This work has led to advances in predicting the success of scientific papers and authors \cite{citepred,predsuc}, found that articles often do not fit into existing disciplinary boundaries \cite{mixedmember,reconcep}, and provided empirical fuel for the debate over interdisciplinary research \cite{interdisc,interdisc_assess,multidisc,socint}. Yet much remains unknown about the nature of the large-scale scientific system that emerges from individuals' intellectual and social incentives. It is especially unclear what features of this system may make it more or less effective at producing insights.

In recent years, network analysis has provided a particularly useful framework for beginning to reveal the structure and evolution of the emergent scientific landscape. The tools of this growing discipline have facilitated greater understanding the roles of specific authors or papers in co-authorship and citation networks. Network measures can predict authors' future collaboration patterns \cite{newman_clustering,moody_netecol}, and can help identify turning points in the literature \cite{chen}. While the fine-scale topology of such networks differs by scientific discipline \cite{newman_main,newman_small}, many display similar global properties. One such commonly shared property is small-world architecture \cite{newman_small,wallace_citation}, which reflects high local clustering within specialty, potentially supporting development and refinement within sub-fields, combined with efficient paths that connect distant areas, providing outlets for innovation and information sharing.

Although co-authorship and citation networks have provided much insight into the properties of the scientific community, their dependence on authors' social network structures makes them an indirect window into the structure of scientific knowledge. \textit{Topic} networks, which  reflect the relations between scientific ideas, offer an opportunity to fill this gap. Surprisingly, the few existing studies of topic networks have largely forgone explicit large-scale analyses in favor of manual inspection of network appearance or node-level trends \cite{novel_keyword,zhang_mapping,co-word,moody_neuro,moody_vfa,frangi}. Yet, the operationalization of science as a set of interconnected ideas provides a unique opportunity to study how research topics are related within and across scientific disciplines, how these topics and their relationships grow and change over time, and how these changes may influence the degree to which scientists engage with the literature.

Here we address these questions in a network of topics covered in \textit{PNAS} since the year 2000. Network nodes reflect specific words or phrases, and network edges reflect the degree of co-occurrence within article abstracts and keyword sections. Using the resultant weighted, undirected network of scientific topics, we address four hypotheses. First, building on findings for co-authorship and citation networks \cite{newman_small,wallace_citation}, we hypothesize that the topic network will demonstrate non-random, small-world structure. Second, based on prior studies that performed latent topic modeling \cite{mixedmember,reconcep}, we hypothesize that the community structure of the network will deviate significantly from disciplinary classifications. Third, as collaboration has crossed national boundaries and broadly increased in recent years \cite{communic,collab}, we hypothesize that over time the network will show greater bridging across topic communities. Fourth, although the benefits of interdisciplinarity for individual papers are debated \cite{interdisc,interdisc_assess,multidisc,socint}, we seek to investigate whether the topic network's interdisciplinarity may be associated with the overall degree of engagement the component literature receives, as measured by \textit{PNAS}'s impact factor.

\begin{figure*}
\centering
\includegraphics[width=1\textwidth]{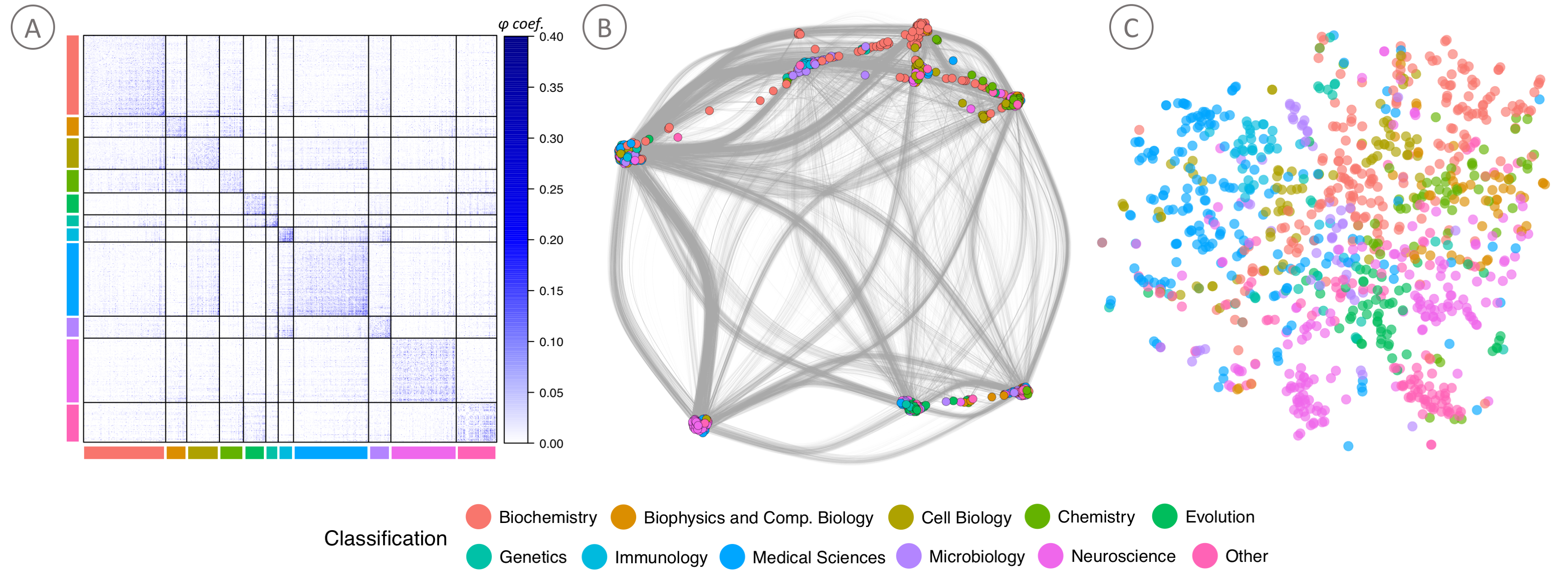}
\caption{\textbf{Architecture of the topic network.} Nodes ($N=1000$) reflect research topics and edges ($E=173,309$) reflect the degree of co-occurrence in abstracts and keyword sections. \textit{(A)} The adjacency matrix sorted by topics' most commonly associated article classification. \textit{(B)} Visualization of the topic network using a distributed recursive graph layout \cite{drl}. \textit{(C)} Visualization of the topic landscape using t-SNE \cite{tsne}, a method that places datapoints on a two-dimensional map based on their similarity. Nodes colored by classification; "other" includes biophysics, developmental biology, ecology, environmental sciences, plant biology, and sustainability science.}
\label{fig:fullim}
\end{figure*}

\section*{Materials and Methods}
\subsection*{Data collection}
We retrieved keywords and abstracts from 65,290 articles published in \textit{PNAS} from the journal's website using an in-house R script, and we used keyword sections to create a list of potential topics to be searched for in the abstracts. This technique was chosen over latent topic modeling, as it reflected scientists' explicit opinions as to the words and phrases that constitute relevant scientific topics, and allowed for the incorporation of multi-word phrases.

\subsection*{Network construction}
We calculated the prevalence of each potential topic by finding the ratio of abstracts or keyword sections containing the topic phrase to the total number of articles written in the time-span of study. Thus, prevalence varied for the full network and the year-specific networks. We used the 1000 most common topics in the given time-span as nodes to construct the network, as this value represented the approximate number at which the least prevalent words occurred often enough to produce meaningful signal. Edges were given by the $\phi$ coefficient for binary association \cite{phi}, representing the degree to which two topics tended to be mentioned in the same abstract. We removed negative correlations, as several statistics for the analysis of signed networks remain difficult to interpret.

We created a dynamic network using a sliding window of $\pm$6 months from a central month. Central months ranged from July, 2000 to May, 2017 such that data from January, 2000 to November, 2017 were included in the analyses. At each window, the 1000 most common topics were used as nodes. We made the choice of 1000 nodes for both the static and dynamic networks because it represented the highest number at which all topics selected in each window would occur more than five times. Thus, higher values would risk uninterpretable noise among low-prevalence topics, and lower values would sacrifice valuable information.

\subsection*{Community detection}
For both the static and the dynamic networks, we performed community detection using an iterative generalized Louvain-like locally greedy algorithm to maximize a common modularity quality function \cite{genlouv}. This technique works by stochastic optimization of the quality index value \textit{Q}, in which nodes are reassigned one by one until no reassignment can improve \textit{Q}, and then by iterating this optimization until convergence to a globally optimal set of community assignments to account for local maxima in the \textit{Q} space. The free parameter, $\gamma$, was selected by maximizing the Jaccard similarity \cite{jaccard} between the community-detection-based partition and the classification-based partition. After averaging over repeated maximizations, this value was determined to be $\gamma=1.2$.

\subsection*{Novel network measures} We defined two novel network measures: deviance and interdisciplinarity. The \textbf{deviance} for a given partition is the degree to which edge weights differ from their expectation under an exponential distribution. It is defined as follows:

$$D_p=\sum_{i \neq j} \left ( w_{ij}-\hat{\beta}_{pb} \right )^2,$$

\noindent where $w_{ij}$ is the observed edge weight in a cell, and $\hat{\beta}_{pb}$ is the expected edge weight for a given block in the partitioned adjacency matrix, as estimated by an exponential model (see \textbf{SI Methods}). The \textbf{interdisciplinarity} is the degree to which a network is well-fit by a small-world structure that does not adhere to a known classification partition. It is defined as follows:

$$\xi= \phi * D_c,$$

\noindent where $\phi$ is the small-world propensity (see \textbf{SI Methods}, \cite{swprop}), and $D_c$ is the deviance under the classification partition.

\section*{Results}
We used data from 65,290 articles published in \textit{PNAS} between January, 2000 and November, 2017 to create a network of research topics. We drew potential topics from the \textit{keywords} section of each article to allow for multi-word phrases. We determined the prevalence of each potential topic by finding the percentage of articles in which the word or phrase was contained in either the abstract or the keywords section. Based on this prevalence score, we identified the 1000 most common topics and represented each as a node in the network (See Methods for details). Edges represented the co-occurrence of topics within abstracts, quantified by the $\phi$ coefficient of association for binary variables \cite{phi} (\textbf{Fig.~\ref{fig:fullim}}). Negative correlations -- comprising roughly 65\% of edges -- were removed to allow for the use of state-of-the-art analysis techniques; these edges had notably lower magnitude and less variability (range: [-0.10,0], interquartile range: 0.004) than the edges that remained (range: [0,0.84], interquartile range: 0.011).

\subsection*{Structure of the topic network}
To understand the structure of the topic network, we calculated measures of interconnectedness (global efficiency) and local clustering (average clustering coefficient); see \textbf{SI Methods} for mathematical definitions. For comparison, we obtained null distributions from 100 random networks with equivalent degree and strength distributions \cite{rubinov}. We observed that the topic network had significantly lower global efficiency ($p<0.01$) and higher average clustering ($p<0.01$) than that observed in the null model, indicating locally dense, non-random connectivity. See \textbf{Table~\ref{tab:s1}} for robustness of results to variations in network size.

\begin{figure}
\centering
\includegraphics[width=1\linewidth]{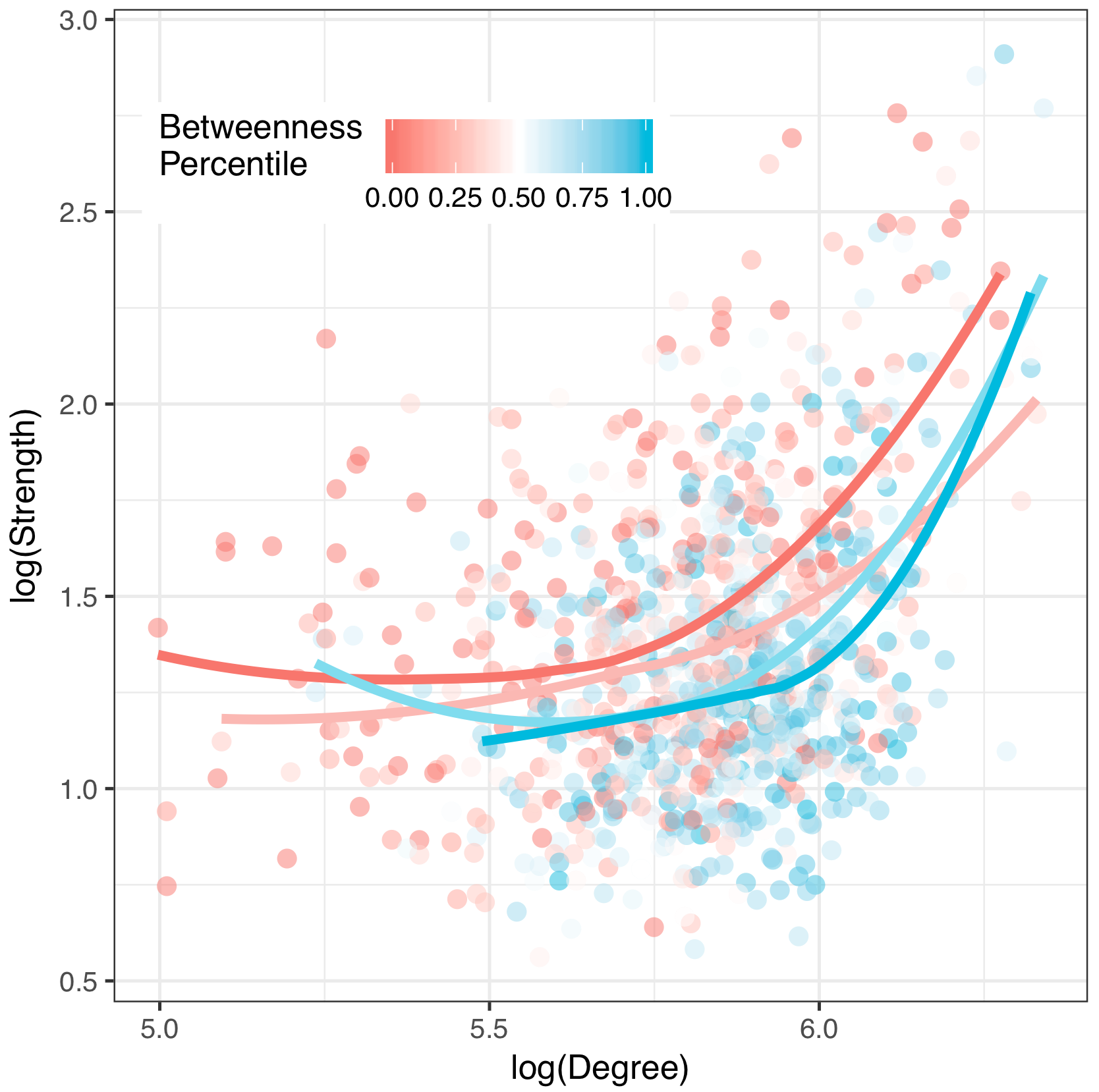}
\caption{\textbf{Relationships among betweenness centrality, degree, and strength of topics.} Topics are colored by their betweenness centrality percentile, from lowest (red) to highest (blue). Lines demonstrate the pattern of degree and strength for nodes in the first quartile (dark red), second quartile (light red), third quartile (light blue), and fourth quartile (dark blue).}
\label{fig:btwstren}
\end{figure}

To probe the local contributions of a topic to this overall structure, we examined each node's general level of connectivity (degree, strength) and its role in bridging disparate regions of the network (betweenness centrality). We observed that betweenness centrality and degree were positively correlated ($\rho=0.30$, $p<0.01$) after accounting for strength, and that betweenness centrality and strength were negatively correlated ($\rho=-0.27$, $p<0.01$) after accounting for degree (\textbf{Fig.~\ref{fig:btwstren}}). These associations indicate that topics with high betweenness centrality tended to be those with many relatively weak connections. Intuitively, this pattern is consistent with the presence of topics that are only occasionally covered but in a wide variety of research areas; topics exemplary of this pattern include \textit{protein expression}, \textit{physiology}, and \textit{molecular mechanism}.

The observed high local clustering and the presence of nodes with high betweenness but low strength could be parsimoniously explained by the principle of small-worldness. To evaluate this possibility, we estimated the small-world propensity (SI Methods, \cite{swprop}); its value was 0.57, significantly higher than would be expected of a random network ($p<0.01$). This result demonstrates that the relationships between topics have small-world properties, with more local clustering than would be expected of a random network and relatively efficient pathways between clusters. The presence of small-worldness then suggests that the landscape of high-quality scientific research is naturally organized into a structure that may be well-suited for advancement within topic clusters and innovation between them.

\subsection*{Community structure of the topic network}
\begin{figure*}
\centering
\includegraphics[width=\linewidth]{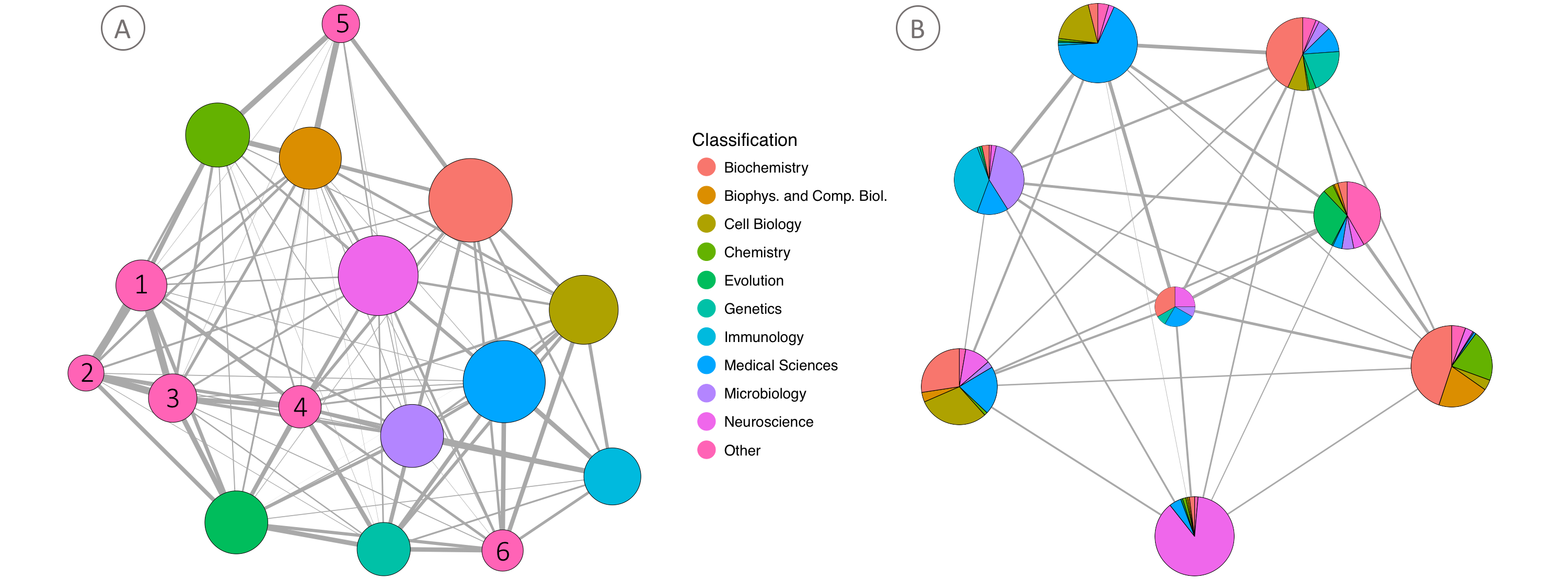}
\caption{\textbf{The topic network's community structure}. Nodes represent topic communities and edges represent the average connection strength between affiliated nodes. \textit{(A)} The graph reflecting communities composed of topics with the same classification. Labels from 1 to 6 refer to environmental sciences, sustainability science, ecology, plant biology, biophysics, and developmental biology, respectively. \textit{(B)} The graph reflecting empirically-identified network communities, derived from modularity maximizations. Nodes show the proportion of topics with each classification in a given community.}
\label{fig:modplot}
\end{figure*}

While the presence of small-worldness in the topic network suggests separation between topic clusters, it remains an open question whether these disparate communities are explained by known disciplinary divisions. To answer this question, we directly compared the communities inherent in the data to the communities implied by 16 disciplinary classifications formally assigned to each \textit{PNAS} publication (see \textbf{Fig.~\ref{fig:fullim}}). To extend the disciplinary classification to individual topics, we assigned each topic the most common classification among articles in which it appeared. We then partitioned topics into communities, where each community was comprised only of topics with a given classification. A visualization of the network according to this classification partition revealed relatively strong connections between communities (\textbf{Fig.~\ref{fig:modplot}A}). 

Next we turned to the problem of identifying a natural partitioning of the topics based solely on the structure of the network, with no knowledge of the disciplinary classifications. We used a Louvain-like locally greedy algorithm \cite{genlouv} to maximize the modularity of the network \cite{newman2006modularity}, thereby obtaining a data-driven partition of the network into communities. The resolution parameter $\gamma$ was determined by maximizing the partition's Jaccard similarity \cite{jaccard} with the partition based on disciplinary classifications in an attempt to optimize comparability (\textbf{Fig.~\ref{fig:gamscrub}}, see Methods for details). The data-driven partition yielded only eight distinct communities, each containing topics from various classifications, with relatively weak connections between communities (\textbf{Fig.~\ref{fig:modplot}B} and \textbf{Table~\ref{tab:s2}}). It can be seen that communities are typically dominated by topics from between one and three classifications. In a subsequent sensitivity analysis, we demonstrate that the partition is robust to the exclusion of negative edge weights (\textbf{Fig.~\ref{fig:negative_cors_final}}).

With the classification partition and data-driven partition in hand, we next sought to quantitatively compare the two. A natural way to formulate this comparison is to calculate the modularity \textit{Q}-value (see Methods for details) for each partition to determine the degree of separation between communities. As the data-driven partition was obtained by optimizing modularity, the magnitude of the increase in \textit{Q} compared to the classification partition demonstrates the degree to which disciplinary classifications do not optimally delineate research topic clusters. We observed that the modularity value was higher in the data-driven partition ($Q=0.37$) than in the classification partition ($Q=0.25$), indicating that the data-driven partition provided a more natural segregation into topic communities. Notably, this effect holds across a range of $\gamma$ values, as the number of communities in the data-driven partition is varied from 8 to 16 (\textbf{Table~\ref{tab:s3}}).

As further confirmation of the data-driven partition's characterization of the community structure, we considered the framework of the weighted stochastic block model (WSBM; \cite{wsbm}), which provides another means of quantifying how well a partition fits the data. Specifically, a WSBM assumes a community structure in which connections within and between communities occur with an expected edge weight. To investigate which partition better characterized the edge weights between and within communities, we fit an exponential model to the edge weights in each within- or between-community block, and calculated the squared difference between the observed edge weights and expected edge weights (see Methods for details). A paired Wilcoxon rank sum test revealed that deviations from the expected weights were significantly higher under the classification partition than the data-driven partition ($p<0.0001$). Notably, this effect also holds across $\gamma$ values (\textbf{Table~\ref{tab:s3}}). Together, these findings indicate that the data-driven partition yields both stronger community separation and greater edge weight consistency within intracommunity and intercommunity blocks.

Critically, not only is the data-driven partition a better fit to the data, but the communities strikingly differ in composition from those defined by classification. In comparison to the classification partition, we observe that the data-driven partition displays lower disciplinarity, as measured by the average proportion of a community's topics that come from its dominant classification. For the classification partition, the disciplinarity value is 1 and for the data-driven partition the value is 0.48, indicating that the average community draws slightly less than half of its topics from its most popular classification. Importantly, the multidisciplinarity of the data-driven partition holds when the number of communities is varied from 8 to 16 (\textbf{Table~\ref{tab:s3}}). Taken together, these findings suggest that the interdisciplinary nature of research published in \textit{PNAS} is underestimated by articles' disciplinary classifications. This implication is reinforced by the fact that only 15\% of articles published since 2000 were given more than one disciplinary classification, whereas 99\% of these articles covered topics from more than one discipline.

\begin{figure}
\centering
\includegraphics[width=1\linewidth]{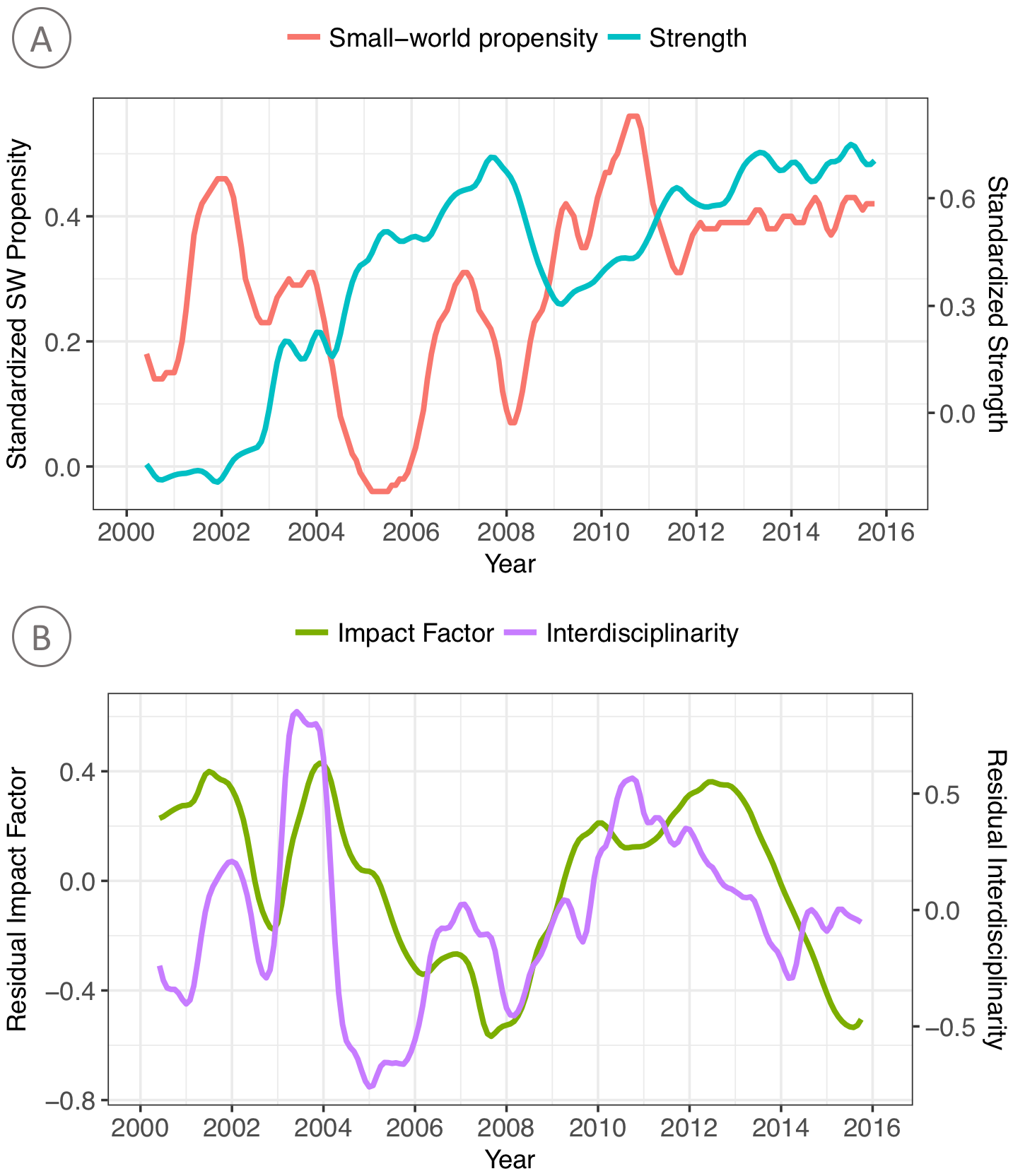}
\caption{\textbf{Temporal features of the dynamic topic network.} \textit{(A)} Temporal trajectories of standardized small-world propensity (orange) and standardized mean edge strength (blue) over time. \textit{(B)} Temporal trajectories of publication-residual impact factor (green) and strength-residual interdisciplinarity $\xi$ (purple).}
\label{fig:impactcors}
\end{figure}

\subsection*{Temporal changes in network structure}
While the static structure of the topic network is important, it does not provide insight into whether and how the landscape of scientific inquiry might change over time. To address this question, we created a dynamic network using a 12-month sliding window with an 11-month overlap over the period from January, 2000 to November, 2017 (see Methods for details). Because some structural change would be expected due to random chance and patterns of journal publication over time, all measures were standardized relative to 100 iterations of a temporal null model where the order of article appearance was permuted uniformly at random. The null trajectories therefore represent change that would occur if topic prevalence and topic associations were stable over the full time period.

We first sought to test our hypothesis that the network would show strengthening connections between and within communities over time, consistent with increasing and changing patterns of collaboration \cite{communic,collab}. We tested for significant temporal changes in strength and small-world propensity by comparing the variance explained by the linear effect of year (R\textsuperscript{2}) to distributions of R\textsuperscript{2} created from the trajectories of the 100 temporal null networks. Average strength ($R^2=0.75$, $p<0.01$) and small-world propensity ($R^2=0.25$, $p=0.01$) both showed significant positive linear trends over time (\textbf{Fig.~\ref{fig:impactcors}A}; see \textbf{Table~\ref{tab:s4}} for consistent trends across network sizes). These results suggest that since 2000, associations between commonly covered scientific topics have grown stronger, and the degree to which these topics demonstrate high clustering and efficient pathways has increased as well.

Next, we sought to investigate whether the network's interdisciplinarity showed a meaningful change over the time period under study. We defined a novel measure of journal interdisciplinarity, $\xi$, given by the product of the network's small-world propensity, $\phi$, and the network's deviance from the classification partition, taken as the mean of the edge deviances described in the previous section (see Methods for details). Therefore, at each temporal window, $\xi$ represents the degree to which the network has small-world structure that is not well-characterized by the assigned topic classifications. We found that classification deviance ($R^2=0.33$, $p<0.01$) and interdisciplinarity ($R^2=0.42$, $p<0.01$) both showed significant positive linear trends. However, because of the strong linear association between strength and time, it is difficult to remove any potential effect of strength on these measures without also eliminating temporal trends.

\subsection*{Implications of interdisciplinarity}
To begin understanding how interdisciplinary research is perceived, we compared the standardized trajectory of interdisciplinarity to the trajectory of \textit{PNAS}'s impact factor. We obtained yearly impact factors from 2000 to 2016 from the Web of Science, and fit a cubic spline to interpolate a smooth monthly trajectory. Interestingly, the number of articles published in a given time window explained 61\% of the variation in impact factor; we therefore only considered the residuals. We calculated the partial correlation between standardized interdisciplinarity and impact factor after accounting for strength \cite{comparing}, and we compared this value to a null correlation distribution, obtained using the set of standardized trajectories drawn from the 100 temporal null networks described previously.

Interdisciplinarity showed a significant, positive partial correlation with impact factor ($r=0.45$, $p=0.03$) (\textbf{Fig.~\ref{fig:impactcors}B}). This result suggests that increases in the interdisciplinarity of scientific topics covered in \textit{PNAS} are associated with increases in the journal's impact. To determine whether this result was driven by only one of the components of $\xi$, we calculated the correlations for small-world propensity and classification deviance separately. Small-worldness alone was not associated with impact factor ($r=0.36$, $p=0.10$), and although deviance did show a significant correlation ($r=0.39$, $p=0.04$), the correlation between interdisciplinarity and impact factor was marginally greater ($p<0.10$) and was more robust to changes in network size (\textbf{Table~\ref{tab:s5}}). These differences indicate that the interdisciplinary small-worldness captured by $\xi$ is likely more associated with external measures of literature engagement than either of its component parts alone.

\section*{Discussion}
Prior analyses of collaboration and citation networks have produced deep insights into the structures and relationships behind the production of scientific research \cite{newman_main,newman_small,wagner_gener,newman_clustering,chen,wallace_citation}. Yet little is known about the network structure of the scientific ideas themselves, or what features of this network might be most effective at facilitating innovation. Here, we set out to characterize the structure of a research topic network, investigate the degree to which topic communities fit into disciplinary classifications, quantify how the landscape of topics is changing over time, and determine whether the network's interdisciplinarity may be related to the degree of engagement that its component research receives.

\subsection*{Structure of the topic network}
We constructed a network of research topics using seventeen years of \textit{PNAS} articles, and found -- unsurprisingly -- that it had features uncharacteristic of a random network. Specifically, the network had significantly higher clustering and lower efficiency than a random network. Interestingly, betweenness centrality of the topics in the network was positively associated with degree, but negatively associated with strength. These results indicate that the network is made up of clustered topic areas that commonly co-occur, and high-degree, low-strength hubs that provide links within and between these clusters. Further supporting this conceptualization is the finding that the network shows a moderate to high degree of small-worldness compared to what would be expected of a random network. Both the graph statistical findings and the small-world classification are consistent with the networks described in studies of co-authorship and citation \cite{newman_small,wallace_citation}, which would be expected to share many features with a network of research topics. 

\subsection*{Community structure of the topic network}
Although community detection was referenced only as a future direction in seminal collaboration network analysis \cite{newman_main}, the modular structure found in the topic network is consistent with the presence of communities in newer research on country-specific collaboration networks in both scientific and nonscientific fields \cite{comscience,commovies}. Comparisons between the community structure that arose from manual disciplinary classification and the empirical community structure inherent to the data were especially revealing. Specifically, compared to the partition arising from disciplinary classifications, the empirical partition showed stronger separation between communities and provided a better fit to the within- and between-community edge weights. Additionally, as opposed to the monodisciplinary classification-specific communities, the empirical communities were found to typically contain an approximately equal balance of topics from two or three distinct disciplines. One notable exception was a community made up almost exclusively of neuroscience topics (\textbf{Fig.~\ref{fig:modplot}B}), potentially reflecting neuroscience's unique status as a field both popular enough to encompass many topics and young enough to remain largely insular.

The superior fit of the empirical communities compared to the classification-specific communities appears to indicate that research published in \textit{PNAS} is more interdisciplinary than the article classifications suggest, an interpretation that is bolstered by the fact that 99\% of articles contained topics from multiple fields while only 15\% of articles were classified under multiple fields. These findings indicate that in spite of the ongoing conflicted discussion regarding the merits and drawbacks of inderdisciplinary research \cite{interdisc,interdisc_assess,multidisc,socint}, researchers publishing in top journals may already be consciously or unconsciously integrating topics and ideas across fields.

\subsection*{Temporal changes in network structure}
Though the structure of the static network yielded valuable insights into the relationships between topics, the production of scientific research is far from static. Therefore, it was of great interest to examine temporal changes in the topic network. At the network scale, while generative evolution has been considered for authorship relationships \cite{wagner_gener,newman_clustering}, the dynamic evolution of large-scale network properties has rarely been examined in the context of authorship or citation \cite{dynauth}. Here we found that both edge strength and small-worldness significantly increased over time. The strengthening of connections between seemingly distant research areas could reflect a convergence of the scientific landscape towards a more interdisciplinary and interconnected network of ideas. This would represent an interesting emergent property of the landscape, potentially arising from individual scientists consciously or unconsciously changing their behavior over time to perform more innovative work.

\subsection*{Implications of interdisciplinarity}
Despite the prevalence of multidisciplinary topic communities and the trend towards stronger cross-field ties, the merits of interdisciplinarity are still widely debated \cite{interdisc,interdisc_assess}. Proponents view interdisciplinary work as being crucial for "address[ing] the great questions of science" \cite{nasint}, while some skeptics instead believe that it too often represents "amateurism and intellectual voyeurism" \cite{davis_grand}. In this study, we defined a novel measure of network interdisciplinarity, $\xi$, and found it to be positively associated with \textit{PNAS}'s impact factor. 

Although this finding only speaks to work of a high caliber, within that context it suggests that bodies of work that are more interdisciplinary in nature may receive more engagement from the scientific community. Yet it remains unclear whether the increased engagement is reflective of the generation of more innovative scientific knowledge, or simply more effective dissemination of the knowledge across fields. In either case, this finding could reflect an important contribution to the discussion of interdisciplinary research, as previous research on the benefits of discipline-spanning has produced mixed results \cite{interdisc_assess}.

\section*{Conclusion}
In this study, we investigated the network characteristics of scientific research topics covered in \textit{PNAS}. The topic network displayed small-world properties and interesting positive degree-betweenness/negative strength-betweenness associations, indicating the presence of tightly connected clusters and high-degree, low-strength hub nodes serving as conceptual bridges. Community detection showed that assigned classifications map poorly onto the underlying clusters, with a data-driven partition revealing the existence of multidisciplinary modules that contained topics from a variety of classifications. By investigating the temporal properties of the network, we found that both strength and small-worldness have been increasing over time. Interestingly, a novel measure of network interdisciplinarity was positively associated with journal impact factor. Overall, this work demonstrates the value of network analysis in gaining insight into the structure of scientific knowledge, paints a picture of the surprisingly integrated nature of scientific ideas, and reveals a potentially important positive relationship between interdisciplinarity and scientific engagement.

\section*{Acknowledgements}
The authors would like to thank Lili Dworkin and Ipek Oguz for advice on data collection and analysis, and Ann E. Sizemore and Richard F. Betzel for helpful feedback on the manuscript. RTS would like to acknowledge support from the National Institute of Neurological Disorders and Stroke (R01 NS085211 \& R01 NS060910). DSB would like to acknowledge support from the John D. and Catherine T. MacArthur Foundation, the Alfred P. Sloan Foundation, and the National Science Foundation CAREER (PHY-1554488). The content is solely the responsibility of the authors and does not necessarily represent the official views of any of the funding agencies.

\subsection*{Data Accessibility}
Data and code can be downloaded at \url{https://www.dropbox.com/sh/p2ksqq8ubvt6haq/AADdlSGfEpX6XL7eA4ua-gl6a?dl=0}. J.D.D. can be reached for questions regarding the code at jdwor@pennmedicine.upenn.edu.

\subsection*{Author Contributions}
J.D.D., R.T.S., and D.S.B. designed research; J.D.D. performed research, wrote the code, and analyzed data; and J.D.D., R.T.S., and D.S.B. wrote the paper.

\bibliographystyle{unsrtnat}
\bibliography{bibfile}

\begin{thebibliography}{53}
\providecommand{\natexlab}[1]{#1}
\providecommand{\url}[1]{\texttt{#1}}
\expandafter\ifx\csname urlstyle\endcsname\relax
  \providecommand{\doi}[1]{doi: #1}\else
  \providecommand{\doi}{doi: \begingroup \urlstyle{rm}\Url}\fi

\bibitem[Wang et~al.(2013)Wang, Song, and Barabási]{citepred}
Dashun Wang, Chaoming Song, and Albert-László Barabási.
\newblock Quantifying {long}-{term} {scientific} {impact}.
\newblock \emph{Science}, 342\penalty0 (6154):\penalty0 127--132, October 2013.
\newblock ISSN 0036-8075, 1095-9203.
\newblock \doi{10.1126/science.1237825}.
\newblock URL
  \url{http://www.sciencemag.org/lookup/doi/10.1126/science.1237825}.

\bibitem[Acuna et~al.(2012)Acuna, Allesina, and Kording]{predsuc}
Daniel~E. Acuna, Stefano Allesina, and Konrad~P. Kording.
\newblock Predicting scientific success: {Future} impact.
\newblock \emph{Nature}, 489\penalty0 (7415):\penalty0 201--202, September
  2012.
\newblock ISSN 0028-0836, 1476-4687.
\newblock \doi{10.1038/489201a}.
\newblock URL \url{http://www.nature.com/articles/489201a}.

\bibitem[Erosheva et~al.(2004)Erosheva, Fienberg, and Lafferty]{mixedmember}
E.~Erosheva, S.~Fienberg, and J.~Lafferty.
\newblock Mixed-membership models of scientific publications.
\newblock \emph{Proceedings of the National Academy of Sciences}, 101\penalty0
  (Supplement 1):\penalty0 5220--5227, April 2004.
\newblock ISSN 0027-8424, 1091-6490.
\newblock \doi{10.1073/pnas.0307760101}.
\newblock URL \url{http://www.pnas.org/cgi/doi/10.1073/pnas.0307760101}.

\bibitem[Airoldi et~al.(2010)Airoldi, Erosheva, Fienberg, Joutard, Love, and
  Shringarpure]{reconcep}
E.~M. Airoldi, E.~A. Erosheva, S.~E. Fienberg, C.~Joutard, T.~Love, and
  S.~Shringarpure.
\newblock Reconceptualizing the classification of {PNAS} articles.
\newblock \emph{Proceedings of the National Academy of Sciences}, 107\penalty0
  (49):\penalty0 20899--20904, December 2010.
\newblock ISSN 0027-8424, 1091-6490.
\newblock \doi{10.1073/pnas.1013452107}.
\newblock URL \url{http://www.pnas.org/cgi/doi/10.1073/pnas.1013452107}.

\bibitem[Rhoten(2004)]{interdisc}
D.~Rhoten.
\newblock {Risks} and {rewards} of an {interdisciplinary} {research} {path}.
\newblock \emph{Science}, 306\penalty0 (5704):\penalty0 2046--2046, December
  2004.
\newblock ISSN 0036-8075, 1095-9203.
\newblock \doi{10.1126/science.1103628}.
\newblock URL \url{http://www.sciencemag.org/cgi/doi/10.1126/science.1103628}.

\bibitem[Jacobs and Frickel(2009)]{interdisc_assess}
Jerry~A. Jacobs and Scott Frickel.
\newblock Interdisciplinarity: {A} {critical} {assessment}.
\newblock \emph{Annual Review of Sociology}, 35\penalty0 (1):\penalty0 43--65,
  August 2009.
\newblock ISSN 0360-0572, 1545-2115.
\newblock \doi{10.1146/annurev-soc-070308-115954}.
\newblock URL
  \url{http://www.annualreviews.org/doi/10.1146/annurev-soc-070308-115954}.

\bibitem[Levitt and Thelwall(2008)]{multidisc}
Jonathan~M. Levitt and Mike Thelwall.
\newblock Is multidisciplinary research more highly cited? {A} macrolevel
  study.
\newblock \emph{Journal of the American Society for Information Science and
  Technology}, 59\penalty0 (12):\penalty0 1973--1984, October 2008.
\newblock ISSN 15322882, 15322890.
\newblock \doi{10.1002/asi.20914}.
\newblock URL \url{http://doi.wiley.com/10.1002/asi.20914}.

\bibitem[Leahey and Moody(2014)]{socint}
Erin Leahey and James Moody.
\newblock Sociological {innovation} through {subfield} {integration}.
\newblock \emph{Social Currents}, 1\penalty0 (3):\penalty0 228--256, October
  2014.
\newblock ISSN 2329-4965, 2329-4973.
\newblock \doi{10.1177/2329496514540131}.
\newblock URL \url{http://journals.sagepub.com/doi/10.1177/2329496514540131}.

\bibitem[Newman(2001{\natexlab{a}})]{newman_clustering}
M.~E.~J. Newman.
\newblock Clustering and preferential attachment in growing networks.
\newblock \emph{Physical Review E}, 64\penalty0 (2), July 2001{\natexlab{a}}.
\newblock ISSN 1063-651X, 1095-3787.
\newblock \doi{10.1103/PhysRevE.64.025102}.
\newblock URL \url{https://link.aps.org/doi/10.1103/PhysRevE.64.025102}.

\bibitem[Borrett et~al.(2014)Borrett, Moody, and Edelmann]{moody_netecol}
Stuart~R. Borrett, James Moody, and Achim Edelmann.
\newblock The rise of {Network} {Ecology}: {Maps} of the topic diversity and
  scientific collaboration.
\newblock \emph{Ecological Modelling}, 293:\penalty0 111--127, December 2014.
\newblock ISSN 03043800.
\newblock \doi{10.1016/j.ecolmodel.2014.02.019}.
\newblock URL
  \url{http://linkinghub.elsevier.com/retrieve/pii/S0304380014001136}.

\bibitem[Chen(2004)]{chen}
C.~Chen.
\newblock Searching for intellectual turning points: {Progressive} knowledge
  domain visualization.
\newblock \emph{Proceedings of the National Academy of Sciences}, 101\penalty0
  (Supplement 1):\penalty0 5303--5310, April 2004.
\newblock ISSN 0027-8424, 1091-6490.
\newblock \doi{10.1073/pnas.0307513100}.
\newblock URL \url{http://www.pnas.org/cgi/doi/10.1073/pnas.0307513100}.

\bibitem[Newman(2004{\natexlab{a}})]{newman_main}
M.~E.~J. Newman.
\newblock Coauthorship networks and patterns of scientific collaboration.
\newblock \emph{Proceedings of the National Academy of Sciences}, 101\penalty0
  (Supplement 1):\penalty0 5200--5205, April 2004{\natexlab{a}}.
\newblock ISSN 0027-8424, 1091-6490.
\newblock \doi{10.1073/pnas.0307545100}.
\newblock URL \url{http://www.pnas.org/cgi/doi/10.1073/pnas.0307545100}.

\bibitem[Newman(2001{\natexlab{b}})]{newman_small}
M.~E.~J. Newman.
\newblock The structure of scientific collaboration networks.
\newblock \emph{Proceedings of the National Academy of Sciences}, 98\penalty0
  (2):\penalty0 404--409, January 2001{\natexlab{b}}.
\newblock ISSN 00278424.
\newblock \doi{10.1073/pnas.021544898}.
\newblock URL \url{\left(}.

\bibitem[Wallace et~al.(2012)Wallace, Larivière, and
  Gingras]{wallace_citation}
Matthew~L. Wallace, Vincent Larivière, and Yves Gingras.
\newblock A {small} {world} of {citations}? {The} {influence} of
  {collaboration} {networks} on {citation} {practices}.
\newblock \emph{PLoS ONE}, 7\penalty0 (3):\penalty0 e33339, March 2012.
\newblock ISSN 1932-6203.
\newblock \doi{10.1371/journal.pone.0033339}.
\newblock URL \url{http://dx.plos.org/10.1371/journal.pone.0033339}.

\bibitem[Radhakrishnan et~al.(2017)Radhakrishnan, Erbis, Isaacs, and
  Kamarthi]{novel_keyword}
Srinivasan Radhakrishnan, Serkan Erbis, Jacqueline~A. Isaacs, and Sagar
  Kamarthi.
\newblock Novel keyword co-occurrence network-based methods to foster
  systematic reviews of scientific literature.
\newblock \emph{PLOS ONE}, 12\penalty0 (3):\penalty0 e0172778, March 2017.
\newblock ISSN 1932-6203.
\newblock \doi{10.1371/journal.pone.0172778}.
\newblock URL \url{http://dx.plos.org/10.1371/journal.pone.0172778}.

\bibitem[Zhang et~al.(2012)Zhang, Xie, Hou, Tu, Xu, Song, Wang, and
  Lu]{zhang_mapping}
Juan Zhang, Jun Xie, Wanli Hou, Xiaochen Tu, Jing Xu, Fujian Song, Zhihong
  Wang, and Zuxun Lu.
\newblock Mapping the {knowledge} {structure} of {research} on {patient}
  {adherence}: {Knowledge} {domain} {visualization} {based} {co}-{word}
  {analysis} and {social} {network} {analysis}.
\newblock \emph{PLoS ONE}, 7\penalty0 (4):\penalty0 e34497, April 2012.
\newblock ISSN 1932-6203.
\newblock \doi{10.1371/journal.pone.0034497}.
\newblock URL \url{http://dx.plos.org/10.1371/journal.pone.0034497}.

\bibitem[Peters and van Raan(1993)]{co-word}
H.P.F. Peters and A.F.J. van Raan.
\newblock Co-word-based science maps of chemical engineering. {Part} {I}:
  {Representations} by direct multidimensional scaling.
\newblock \emph{Research Policy}, 22\penalty0 (1):\penalty0 23--45, February
  1993.
\newblock ISSN 00487333.
\newblock \doi{10.1016/0048-7333(93)90031-C}.
\newblock URL
  \url{http://linkinghub.elsevier.com/retrieve/pii/004873339390031C}.

\bibitem[Beam et~al.(2014)Beam, Appelbaum, Jack, Moody, and
  Huettel]{moody_neuro}
Elizabeth Beam, L.~Gregory Appelbaum, Jordynn Jack, James Moody, and Scott~A.
  Huettel.
\newblock Mapping the {semantic} {structure} of {cognitive} {neuroscience}.
\newblock \emph{Journal of Cognitive Neuroscience}, 26\penalty0 (9):\penalty0
  1949--1965, September 2014.
\newblock ISSN 0898-929X, 1530-8898.
\newblock \doi{10.1162/jocn\_a\_00604}.
\newblock URL \url{http://www.mitpressjournals.org/doi/10.1162/jocn_a_00604}.

\bibitem[Moody and Light(2006)]{moody_vfa}
James Moody and Ryan Light.
\newblock A view from above: {The} evolving sociological landscape.
\newblock \emph{The American Sociologist}, 37\penalty0 (2):\penalty0 67--86,
  June 2006.
\newblock ISSN 0003-1232, 1936-4784.
\newblock \doi{10.1007/s12108-006-1006-8}.
\newblock URL \url{http://link.springer.com/10.1007/s12108-006-1006-8}.

\bibitem[Frangi et~al.(2016)Frangi, Taylor, and Gooya]{frangi}
Alejandro~F. Frangi, Zeike~A. Taylor, and Ali Gooya.
\newblock Precision {Imaging}: more descriptive, predictive and integrative
  imaging.
\newblock \emph{Medical Image Analysis}, 33:\penalty0 27--32, October 2016.
\newblock ISSN 13618415.
\newblock \doi{10.1016/j.media.2016.06.024}.
\newblock URL
  \url{http://linkinghub.elsevier.com/retrieve/pii/S1361841516301049}.

\bibitem[Hoekman et~al.(2010)Hoekman, Frenken, and Tijssen]{communic}
Jarno Hoekman, Koen Frenken, and Robert~J.W. Tijssen.
\newblock Research collaboration at a distance: {Changing} spatial patterns of
  scientific collaboration within {Europe}.
\newblock \emph{Research Policy}, 39\penalty0 (5):\penalty0 662--673, June
  2010.
\newblock ISSN 00487333.
\newblock \doi{10.1016/j.respol.2010.01.012}.
\newblock URL
  \url{http://linkinghub.elsevier.com/retrieve/pii/S0048733310000260}.

\bibitem[Huang et~al.(2008)Huang, Zhuang, Li, and Giles]{collab}
Jian Huang, Ziming Zhuang, Jia Li, and C.~Lee Giles.
\newblock Collaboration over time: characterizing and modeling network
  evolution.
\newblock page 107. ACM Press, 2008.
\newblock \doi{10.1145/1341531.1341548}.
\newblock URL \url{http://portal.acm.org/citation.cfm?doid=1341531.1341548}.

\bibitem[Martin et~al.(2008)Martin, Brown, Klavans, and Boyack]{drl}
S.~Martin, W.~M. Brown, R.~Klavans, and K.~W. Boyack.
\newblock Drl: Distributed recursive (graph) layout.
\newblock \emph{SAND Reports}, 2936:\penalty0 1--10, 2008.

\bibitem[van~der Maaten and Hinton(2008)]{tsne}
L.J.P. van~der Maaten and G.E. Hinton.
\newblock Visualizing high-dimensional data using t-sne.
\newblock \emph{Journal of Machine Learning Research}, 9:\penalty0 2579--2605,
  01 2008.

\bibitem[Davenport and El-Sanhurry(1991)]{phi}
Ernest~C. Davenport and Nader~A. El-Sanhurry.
\newblock Phi/{Phimax}: {Review} and {synthesis}.
\newblock \emph{Educational and Psychological Measurement}, 51\penalty0
  (4):\penalty0 821--828, December 1991.
\newblock ISSN 0013-1644, 1552-3888.
\newblock \doi{10.1177/001316449105100403}.
\newblock URL \url{http://journals.sagepub.com/doi/10.1177/001316449105100403}.

\bibitem[Jeub et~al.(2011)Jeub, Bazzi, Jutla, and Mucha]{genlouv}
L.G.S. Jeub, M.~Bazzi, I.S. Jutla, and P.J. Mucha.
\newblock A generalized {Louvain} method for community detection implemented in
  {MATLAB}, 2011.
\newblock URL \url{http://netwiki.amath.unc.edu/GenLouvain}.

\bibitem[Steen et~al.(2011)Steen, Hayasaka, Joyce, and Laurienti]{jaccard}
Matthew Steen, Satoru Hayasaka, Karen Joyce, and Paul Laurienti.
\newblock Assessing the consistency of community structure in complex networks.
\newblock \emph{Physical Review E}, 84\penalty0 (1), July 2011.
\newblock ISSN 1539-3755, 1550-2376.
\newblock \doi{10.1103/PhysRevE.84.016111}.
\newblock URL \url{https://link.aps.org/doi/10.1103/PhysRevE.84.016111}.

\bibitem[Muldoon et~al.(2016)Muldoon, Bridgeford, and Bassett]{swprop}
Sarah~Feldt Muldoon, Eric~W. Bridgeford, and Danielle~S. Bassett.
\newblock Small-{world} {propensity} and {weighted} {brain} {networks}.
\newblock \emph{Scientific Reports}, 6\penalty0 (1), April 2016.
\newblock ISSN 2045-2322.
\newblock \doi{10.1038/srep22057}.
\newblock URL \url{http://www.nature.com/articles/srep22057}.

\bibitem[Rubinov and Sporns(2011)]{rubinov}
Mikail Rubinov and Olaf Sporns.
\newblock Weight-conserving characterization of complex functional brain
  networks.
\newblock \emph{NeuroImage}, 56\penalty0 (4):\penalty0 2068--2079, June 2011.
\newblock ISSN 10538119.
\newblock \doi{10.1016/j.neuroimage.2011.03.069}.
\newblock URL
  \url{http://linkinghub.elsevier.com/retrieve/pii/S105381191100348X}.

\bibitem[Newman(2006)]{newman2006modularity}
M~E~J Newman.
\newblock Modularity and community structure in networks.
\newblock \emph{Proc. Natl. Acad. Sci. U.S.A.}, 103\penalty0 (23):\penalty0
  8577--–8582, 2006.

\bibitem[Aicher et~al.(2015)Aicher, Jacobs, and Clauset]{wsbm}
C.~Aicher, A.~Z. Jacobs, and A.~Clauset.
\newblock Learning latent block structure in weighted networks.
\newblock \emph{Journal of Complex Networks}, 3\penalty0 (2):\penalty0
  221--248, June 2015.
\newblock ISSN 2051-1310, 2051-1329.
\newblock \doi{10.1093/comnet/cnu026}.
\newblock URL
  \url{https://academic.oup.com/comnet/article-lookup/doi/10.1093/comnet/cnu026}.

\bibitem[van Wijk et~al.(2010)van Wijk, Stam, and Daffertshofer]{comparing}
Bernadette C.~M. van Wijk, Cornelis~J. Stam, and Andreas Daffertshofer.
\newblock Comparing {brain} {networks} of {different} {size} and {connectivity}
  {density} {using} {graph} {theory}.
\newblock \emph{PLoS ONE}, 5\penalty0 (10):\penalty0 e13701, October 2010.
\newblock ISSN 1932-6203.
\newblock \doi{10.1371/journal.pone.0013701}.
\newblock URL \url{http://dx.plos.org/10.1371/journal.pone.0013701}.

\bibitem[Wagner and Leydesdorff(2005)]{wagner_gener}
Caroline~S. Wagner and Loet Leydesdorff.
\newblock Network structure, self-organization, and the growth of international
  collaboration in science.
\newblock \emph{Research Policy}, 34\penalty0 (10):\penalty0 1608--1618,
  December 2005.
\newblock ISSN 00487333.
\newblock \doi{10.1016/j.respol.2005.08.002}.
\newblock URL
  \url{http://linkinghub.elsevier.com/retrieve/pii/S0048733305001745}.

\bibitem[Lužar et~al.(2014)Lužar, Levnajić, Povh, and Perc]{comscience}
Borut Lužar, Zoran Levnajić, Janez Povh, and Matjaž Perc.
\newblock Community {structure} and the {evolution} of {interdisciplinarity} in
  {Slovenia}'s {scientific} {collaboration} {network}.
\newblock \emph{PLoS ONE}, 9\penalty0 (4):\penalty0 e94429, April 2014.
\newblock ISSN 1932-6203.
\newblock \doi{10.1371/journal.pone.0094429}.
\newblock URL \url{http://dx.plos.org/10.1371/journal.pone.0094429}.

\bibitem[Battiston et~al.(2016)Battiston, Iacovacci, Nicosia, Bianconi, and
  Latora]{commovies}
Federico Battiston, Jacopo Iacovacci, Vincenzo Nicosia, Ginestra Bianconi, and
  Vito Latora.
\newblock Emergence of {multiplex} {communities} in {collaboration} {networks}.
\newblock \emph{PLOS ONE}, 11\penalty0 (1):\penalty0 e0147451, January 2016.
\newblock ISSN 1932-6203.
\newblock \doi{10.1371/journal.pone.0147451}.
\newblock URL \url{http://dx.plos.org/10.1371/journal.pone.0147451}.

\bibitem[Kronegger et~al.(2011)Kronegger, Ferligoj, and Doreian]{dynauth}
Luka Kronegger, Anuška Ferligoj, and Patrick Doreian.
\newblock On the dynamics of national scientific systems.
\newblock \emph{Quality \& Quantity}, 45\penalty0 (5):\penalty0 989--1015,
  August 2011.
\newblock ISSN 0033-5177, 1573-7845.
\newblock \doi{10.1007/s11135-011-9484-3}.
\newblock URL \url{http://link.springer.com/10.1007/s11135-011-9484-3}.

\bibitem[nas(2004)]{nasint}
\emph{Facilitating {interdisciplinary} {research}}.
\newblock National Academies Press, Washington, D.C., April 2004.
\newblock ISBN 978-0-309-09435-1.
\newblock URL \url{http://www.nap.edu/catalog/11153}.
\newblock DOI: 10.17226/11153.

\bibitem[Davis(2007)]{davis_grand}
Lennard~J Davis.
\newblock A grand unified theory of interdisciplinarity.
\newblock \emph{Chronicles of Higher Education}, 53\penalty0 (40):\penalty0 B9,
  2007.

\bibitem[Bordons et~al.(2002)Bordons, Fern\'{a}ndez, and G\'{o}mez]{imfac1}
M.~Bordons, M.~T. Fern\'{a}ndez, and I.~G\'{o}mez.
\newblock Advantages and limitations in the use of impact factor measures for
  the assessment of research performance.
\newblock \emph{Scientometrics}, 53:\penalty0 195--206, 02 2002.

\bibitem[Kurmis(2003)]{imfac2}
Andrew~P. Kurmis.
\newblock Understanding the limitations of the journal impact factor.
\newblock \emph{The Journal of Bone and Joint Surgery. American Volume},
  85-A\penalty0 (12):\penalty0 2449--2454, December 2003.
\newblock ISSN 0021-9355.

\bibitem[Chew et~al.(2007)Chew, Villanueva, and Van Der~Weyden]{imfac3}
M.~Chew, E.~V Villanueva, and M.~B Van Der~Weyden.
\newblock Life and times of the impact factor: retrospective analysis of trends
  for seven medical journals (1994-2005) and their {Editors}' views.
\newblock \emph{Journal of the Royal Society of Medicine}, 100\penalty0
  (3):\penalty0 142--150, March 2007.
\newblock ISSN 0141-0768, 0141-0768.
\newblock \doi{10.1258/jrsm.100.3.142}.
\newblock URL \url{http://www.jrsm.org/cgi/doi/10.1258/jrsm.100.3.142}.

\bibitem[rst(2015)]{rstat}
R: {A} {language} and {environment} for {statistical} {computing}, 2015.

\bibitem[Csardi and Nepusz(2006)]{igraph}
Gabor Csardi and Tamas Nepusz.
\newblock The igraph software package for complex network research.
\newblock \emph{InterJournal}, Complex Systems:\penalty0 1695, 2006.
\newblock URL \url{http://igraph.org}.

\bibitem[Rubinov and Sporns(2010)]{bct}
Mikail Rubinov and Olaf Sporns.
\newblock Complex network measures of brain connectivity: {Uses} and
  interpretations.
\newblock \emph{NeuroImage}, 52\penalty0 (3):\penalty0 1059--1069, September
  2010.
\newblock ISSN 10538119.
\newblock \doi{10.1016/j.neuroimage.2009.10.003}.
\newblock URL
  \url{http://linkinghub.elsevier.com/retrieve/pii/S105381190901074X}.

\bibitem[{Harrell Jr} et~al.(2016){Harrell Jr}, with contributions~from
  Charles~Dupont, and many others.]{hmisc}
Frank~E {Harrell Jr}, with contributions~from Charles~Dupont, and many others.
\newblock \emph{Hmisc: Harrell Miscellaneous}, 2016.
\newblock URL \url{https://CRAN.R-project.org/package=Hmisc}.
\newblock R package version 4.0-2.

\bibitem[Freeman(1978)]{freeman_centrality_1978}
Linton~C. Freeman.
\newblock Centrality in social networks conceptual clarification.
\newblock \emph{Social Networks}, 1\penalty0 (3):\penalty0 215--239, January
  1978.
\newblock ISSN 03788733.
\newblock \doi{10.1016/0378-8733(78)90021-7}.
\newblock URL
  \url{http://linkinghub.elsevier.com/retrieve/pii/0378873378900217}.

\bibitem[Barrat et~al.(2004)Barrat, Barthelemy, Pastor-Satorras, and
  Vespignani]{barrat}
A.~Barrat, M.~Barthelemy, R.~Pastor-Satorras, and A.~Vespignani.
\newblock The architecture of complex weighted networks.
\newblock \emph{Proceedings of the National Academy of Sciences}, 101\penalty0
  (11):\penalty0 3747--3752, March 2004.
\newblock ISSN 0027-8424, 1091-6490.
\newblock \doi{10.1073/pnas.0400087101}.
\newblock URL \url{http://www.pnas.org/cgi/doi/10.1073/pnas.0400087101}.

\bibitem[Latora and Marchiori(2001)]{effic}
Vito Latora and Massimo Marchiori.
\newblock Efficient {behavior} of {small}-{world} {networks}.
\newblock \emph{Physical Review Letters}, 87\penalty0 (19), October 2001.
\newblock ISSN 0031-9007, 1079-7114.
\newblock \doi{10.1103/PhysRevLett.87.198701}.
\newblock URL \url{https://link.aps.org/doi/10.1103/PhysRevLett.87.198701}.

\bibitem[Watts and Strogatz(1998)]{watts_small}
D.~J. Watts and S.~H. Strogatz.
\newblock Collective dynamics of 'small-world' networks.
\newblock \emph{Nature}, 393\penalty0 (6684):\penalty0 440--442, June 1998.
\newblock ISSN 0028-0836.
\newblock \doi{10.1038/30918}.

\bibitem[Newman(2004{\natexlab{b}})]{modul}
M.~E.~J. Newman.
\newblock Analysis of weighted networks.
\newblock \emph{Physical Review E}, 70\penalty0 (5), November
  2004{\natexlab{b}}.
\newblock ISSN 1539-3755, 1550-2376.
\newblock \doi{10.1103/PhysRevE.70.056131}.
\newblock URL \url{https://link.aps.org/doi/10.1103/PhysRevE.70.056131}.

\bibitem[Gómez et~al.(2009)Gómez, Jensen, and Arenas]{signed}
Sergio Gómez, Pablo Jensen, and Alex Arenas.
\newblock Analysis of community structure in networks of correlated data.
\newblock \emph{Physical Review E}, 80\penalty0 (1), July 2009.
\newblock ISSN 1539-3755, 1550-2376.
\newblock \doi{10.1103/PhysRevE.80.016114}.
\newblock URL \url{https://link.aps.org/doi/10.1103/PhysRevE.80.016114}.

\bibitem[Good et~al.(2010)Good, de~Montjoye, and Clauset]{neardeg}
Benjamin~H. Good, Yves-Alexandre de~Montjoye, and Aaron Clauset.
\newblock Performance of modularity maximization in practical contexts.
\newblock \emph{Physical Review E}, 81\penalty0 (4), April 2010.
\newblock ISSN 1539-3755, 1550-2376.
\newblock \doi{10.1103/PhysRevE.81.046106}.
\newblock URL \url{https://link.aps.org/doi/10.1103/PhysRevE.81.046106}.

\bibitem[Wang and Wong(1987)]{stoch}
Yuchung~J. Wang and George~Y. Wong.
\newblock Stochastic {blockmodels} for {directed} {graphs}.
\newblock \emph{Journal of the American Statistical Association}, 82\penalty0
  (397):\penalty0 8--19, March 1987.
\newblock ISSN 0162-1459, 1537-274X.
\newblock \doi{10.1080/01621459.1987.10478385}.
\newblock URL
  \url{http://www.tandfonline.com/doi/abs/10.1080/01621459.1987.10478385}.

\end{thebibliography}

\clearpage

\appendix
\newcommand{\hbAppendixPrefix}{S}
\renewcommand{\thefigure}{\hbAppendixPrefix\arabic{figure}}
\setcounter{figure}{0}
\renewcommand{\thetable}{\hbAppendixPrefix\arabic{table}} 
\setcounter{table}{0}

\section*{Supporting Information}
\subsection*{SI Text}
\subsubsection*{Limitations}
Validity and generalizability of the findings presented in this paper are limited by a few methodological considerations. First, seminal work has shown that fields of study differ significantly in the structures of their authorship and citation networks \cite{newman_main,newman_small}. Therefore it is conceivable that journals may also have meaningful differences in the structure and correlates of topic networks. If this is the case, using journal-specific data may limit the degree to which the findings in this paper can be generalized to scientific research more broadly. Future work could expand the data source to include several top-tier journals within and across fields.

Additionally, the restriction of the dataset to keyword sections and abstracts may ignore potential information contained in introduction and discussion sections. However, it is plausible that topics mentioned in introduction and discussion areas may not be an accurate reflection of the topics truly covered in a given article, unlike those mentioned in the abstract and keyword sections. Future work could examine this assumption directly by implementing a manual rating system.

Finally, impact factor is widely considered to be an imperfect measure of scientific engagement with published research. Although warnings against impact factor's use often highlight its inability to facilitate valid comparisons between journals in different fields or different countries \cite{imfac1,imfac2}, within-journal changes over time are also incomplete and potentially subject to manipulation through editorial policies \cite{imfac3}. Future work could consider associations between network structure and other measures of scientific engagement and journal quality.

\subsection*{SI Methods}
\subsubsection*{Data analysis}
Topic networks and temporal null networks were created, visualized, and analyzed in the R statistical environment \cite{rstat} using the iGraph package \cite{igraph}. Benchmark random networks were generated using the Brain Connectivity Toolbox in MATLAB \cite{bct}, but were analyzed using the iGraph package. Community detection was carried out in MATLAB using the GenLouvain toolbox \cite{genlouv}. Correlation coefficients and probability values were obtained using the Hmisc package in R \cite{hmisc}. 

\subsubsection*{Network Measures} Here we provide a brief description of the network measures used in this study.

The \textbf{degree} of a node is the number of edges, regardless of weight, connected to the node \cite{bct}. Degree then represents one aspect of the node's importance, measured by the number of neighbors it has in the network. It is defined as follows:

$$k_i=\sum_{j\in N} a_{ij},$$

\noindent where $N$ is the set of all nodes in the network, and $a_{ij}$ is 1 if nodes $i$ and $j$ are connected by an edge and 0 if not.

\bigskip The \textbf{strength} of a node is the sum of the weights of all edges connected to the node \cite{bct}. This measure is similar to degree in that it sums a node's connecting edges, but strength additionally allows for edges of varying weights. It is defined as follows:

$$s_i=\sum_{j\in N} w_{ij},$$

\noindent where $w_{ij}$ is the weight of the edge between nodes $i$ and $j$ if they are connected and 0 if not.

\bigskip The \textbf{betweenness centrality} of a node is the proportion of all shortest paths within the network that pass through the given node \cite{freeman_centrality_1978}. Betweenness centrality represents the degree to which a specific node functions as a bridge between nodes in disparate parts of the network. It is defined as follows:

$$b_i=\frac{1}{(n-1)(n-2)}\sum_{h,j\in N; h\neq j, h\neq i, j\neq i} \frac{\rho_{hj}^{(i)}}{\rho_{hj}},$$

\noindent where $\rho_{hj}$ is the number of shortest weighted paths between $h$ and $j$, $\rho_{hj}^{(i)}$ is the number of shortest weighted paths between $h$ and $j$ that pass through node $i$, and $n$ is the number of nodes in the graph.

\bigskip The \textbf{clustering coefficient} of a node can be defined as the probability that two of its adjacent nodes are connected to each other. A node's clustering coefficient then represents the amount of interconnectedness in a node's local neighborhood. The version used in the current study is a measure of transitivity, as given by Barrat \cite{barrat}. It is defined as follows:

$$c_i^{w}=\frac{1}{s_i(k_i-1)}\sum_{h,j\in N} \frac{(w_{ij}+w_{ih})}{2}a_{ij}a_{ih}a_{hj}.$$

\bigskip The \textbf{global efficiency} of a network can be defined as the average inverse shortest path length between any two nodes \cite{effic}. Global efficiency is often thought of as representing the amount of integration within and between disparate parts of the network. It is defined as follows:

$$E^w=\frac{1}{n}\sum_{i\in N} \frac{\sum_{j\in N; j\neq i}d_{ij}^{-1}}{n-1},$$

\noindent where $d_{ij}$ is the shortest weighted path length between node $i$ and node $j$.

\bigskip The \textbf{path length} of a network is the average shortest path length between all node pairs \cite{watts_small}. In many graphs, path length is inversely correlated with global efficiency, and is therefore often interpreted as representing an alternative measure of network integration. A version of the path length for a weighted network is as follows:

$$L=\frac{1}{n(n-1)}\sum_{i\neq j} d_{ij}.$$

\bigskip The \textbf{modularity} of a network intuitively represents the degree of separation between nodes in different groups \cite{modul}. It quantifies how well the network can be separated into non-overlapping communities, with many within-group connections and few between-group connections. For a network containing only positive weights, the modularity can be defined as follows:

$$Q^w=\frac{1}{l^w}\sum_{i,j\in N} \left [ w_{ij} - \frac{s_is_j}{l^w} \right ]\delta_{m_im_j},$$

and for a signed network, the modularity can be defined as follows \cite{signed}:

$$Q_s^w=\frac{1}{l_+^w + l_-^w}\sum_{i,j\in N} \left [ w_{ij} - \frac{s_i^+ s_j^+}{l_+^w} + \frac{s_i^- s_j^-}{l_-^w} \right ]\delta_{m_im_j},$$

\noindent where $l^w$ is the sum of all of the weights in the network, $l_+^w$ is the sum of all of the positive weights in the network, $l_-^w$ is the sum of all of the negative weights in the network, $s_i^+$ is the strength of a node's positive edges, $s_i^-$ is the strength of a node's negative edges, and $\delta_{m_im_j}$ is 1 if $i=j$ and 0 otherwise. Here, we addressed the issue of near degeneracy \cite{neardeg} in the modularity landscape by using 100 iterations of a Louvain-like locally greedy algorithm to maximize the modularity quality function \cite{genlouv}, and we report the consensus partition over those iterations.

\bigskip The \textbf{small-world propensity} is the degree to which a network shows similar clustering to that of a lattice network, and similar average path length to that of a random network \cite{swprop}. This metric is similar to the commonly used small-world index, $\sigma$ \cite{watts_small}, but has been shown to be unbiased even in the context of networks with varying densities. Both measures broadly represent how well a network can be characterized as having both disparate clusters and high levels of between-cluster integration. Small-world propensity is defined as follows:

$$\phi = 1 - \sqrt{\frac{\Delta_C^2 + \Delta_L^2}{2}},$$

\noindent where

$$\Delta_C = \frac{C_{lattice}-C_{observed}}{C_{lattice}-C_{random}},$$

\noindent and

$$\Delta_L = \frac{L_{observed}-L_{random}}{L_{lattice}-L_{random}},$$

\noindent with $C$ representing the network clustering coefficient, defined as the average node-specific $c_i^w$ values.

\bigskip The \textbf{stochastic block model} assumes a community structure in which between- and within-group connections occur with a specific probability (in the unweighted case) or an expected edge weight (in the weighted case). Unlike modularity, which characterizes a community structure with many (strong) connections within groups and few (weak) connections between groups, the stochastic block model characterizes a community structure with consistent connection patterns within and between groups. For the unweighted case \cite{stoch}, it is defined as follows:

$$P_{g,B}(A) = \prod_{i\neq j} B_{g_i g_j}^{a_{ij}} \left ( 1-B_{g_i g_j} \right ) ^{\left ( 1- a_{ij} \right )},$$

\noindent where $g\in {1,...,K}^n$ is a vector of community memberships, assuming $K$ distinct communities, and $B\in[0,1]^{KxK}$ is a matrix of community-wise edge probabilities.

For the exponential weighted framework used in the current study \cite{wsbm}, the model is defined as follows:

$$P_{g,\Lambda}(A) = \prod_{i\neq j} \Lambda_{g_i g_j} e^{-\Lambda_{g_i g_j} w_{ij}},$$

\noindent where $\Lambda \in [0,\infty)^{KxK}$ is a matrix of community-wise rate parameters.

\begin{table*}[ht]
\centering
\small
\begin{tabular}{llll}
Network Measure & N = 950 & N = 1000 & N = 1050 \\
\midrule
Small-world propensity ($\phi$) & 0.59** & 0.58** & 0.57** \\
Betweenness - degree correlation ($r$) & 0.28** & 0.30** & 0.31** \\
Betweenness - strength correlation ($r$) & -0.31** & -0.27** & -0.23**\\
\bottomrule
\end{tabular}
\caption{Effect of network size on the results for the full network. \textmd{Rows represent different network-level measures reported in the full text, columns represent their values and statistical significance for different choices of network size. Note: * = $p < 0.05$, ** = $p < 0.01$.}}
\label{tab:s1}
\end{table*}

\begin{table*}[ht]
\centering
\small
\begin{tabular}{lllll}
Group & \# of topics & Primary topic classification & Secondary topic classification & Tertiary topic classification \\
\midrule
1 & 206 & Biochemistry (49\%) & Chemistry (21\%)  & Biophysics and Comp. Biol (18\%)\\
2 & 178 & Medical Sciences (68\%) & Cell Biology (19\%) & Developmental Biology (4\%) \\
3 & 175 & Evolution (29\%) & Environmental Sciences (15\%) & Ecology (13\%) \\
4 & 141 & Neuroscience (89\%) & Medical Sciences (4\%) & Biochemistry (2\%) \\
5 & 117 & Biochemistry (43\%) & Genetics (20\%) & Medical Sciences (14\%) \\
6 & 92 & Microbiology (40\%) & Immunology (37\%) & Medical Sciences (14\%) \\
7 & 78 & Biochemistry (32\%) & Cell Biology (28\%) & Medical Sciences (19\%) \\
8 & 13 & Biochemistry (38\%) & Medical Sciences (23\%) & Neuroscience (23\%) \\
\bottomrule
\end{tabular}
\caption{Classification composition of empirically obtained topic communities. \textmd{Rows represent the eight communities, and columns give the three most common classifications for the topics contained within each community.}}
\label{tab:s2}
\end{table*}

\begin{table*}[ht]
\centering
\small
\begin{tabular}{llll}
\begin{tabular}{@{}c@{}}Number of\\communities\end{tabular} & Modularity (Q) & Disciplinarity & \begin{tabular}{@{}c@{}}P-value for test of\\ WSBM deviance\end{tabular}\\
\midrule
9 & 0.36 & 0.52 & $<0.0001$\\
10 & 0.35 & 0.53 & $<0.0001$\\
11 & 0.35 & 0.53 & $<0.0001$\\
12 & 0.34 & 0.53 & $<0.0001$\\
13 & 0.34 & 0.54 & $<0.0001$\\
14 & 0.34 & 0.55 & $<0.0001$\\
15 & 0.33 & 0.56 & $<0.0001$\\
16 & 0.33 & 0.56 & $<0.0001$\\
\bottomrule
\end{tabular}
\caption{Effect of the number of communities on features of the empirical partition. \textmd{Rows represent partitions with between 9 and 16 communities. Columns represent the degree to which the partitions demonstrate modular structure, contain disciplinary communities, and better explain edge weights compared to the classification partition.}}
\label{tab:s3}
\end{table*}

\begin{table*}[ht]
\centering
\small
\begin{tabular}{llll}
Network Measure & N = 950 & N = 1000 & N = 1050 \\
\midrule
Strength by time ($R^2$) & 0.77** & 0.75** & 0.74** \\
Interdisciplinarity by time ($R^2$) & 0.41** & 0.42** & 0.40**\\
Small-world propensity by time ($R^2$) & 0.26* & 0.25* & 0.20 \\
Classification deviance by time ($R^2$) & 0.31** & 0.33** & 0.33**\\
\bottomrule
\end{tabular}
\caption{Effect of network size on the linear trajectories of the temporal network. \textmd{Rows represent the variances explained by time for various measures of the temporal network. Columns represent their values and statistical significance for different choices of network size. Note: * = $p < 0.05$, ** = $p < 0.01$.}}
\label{tab:s4}
\end{table*}

\begin{table*}[ht]
\centering
\small
\begin{tabular}{llll}
Network Measure & N = 950 & N = 1000 & N = 1050 \\
\midrule
Interdisciplinarity - impact factor correlation ($r$) & 0.44* & 0.45* & 0.45*\\
Small-world propensity - impact factor correlation ($r$) & 0.35 & 0.36 & 0.35 \\
Classification deviance - impact factor correlation ($r$) & 0.37 & 0.39* & 0.38\\
\bottomrule
\end{tabular}
\caption{Effect of network size on the impact factor correlations of the temporal network. \textmd{Rows represent the correlations with \textit{PNAS}'s impact factor for various measures of the temporal network. Columns represent their values and statistical significance for different choices of network size. Note: * = $p < 0.05$, ** = $p < 0.01$.}}
\label{tab:s5}
\end{table*}

\clearpage

\begin{figure}[ht]
\centering
\includegraphics[width=1\linewidth]{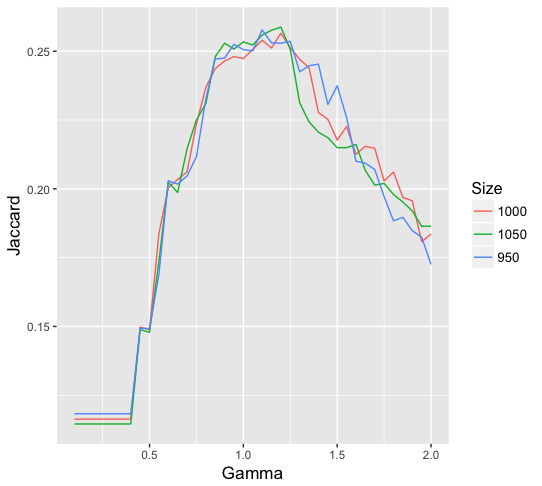}
\caption{Visualization of the Jaccard similarity between the empirical community structure and the assigned topic classifications. Jaccard similarities are plotted for a range of $\gamma$ values, demonstrating the procedure for optimizing Jaccard similarity over $\gamma$ that was used when performing community detection. These values are shown for three different choices of network size.}
\label{fig:gamscrub}
\end{figure}

\begin{figure}[ht]
\centering
\includegraphics[width=1\linewidth]{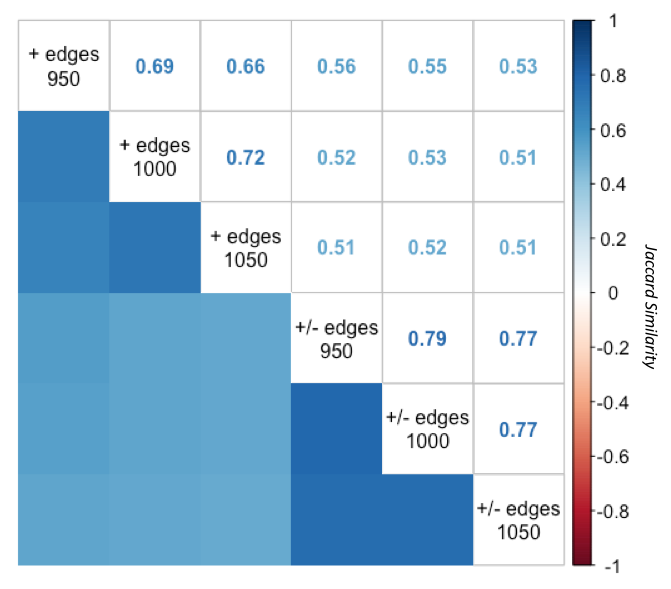}
\caption{Visualization of the consistency, using Jaccard similarity, of the empirical community structure both (i) across sizes, and (ii) with or without negative edge weights. Community structure was consistent across sizes, and was reasonably consistent between positive weighted networks, and positive-and-negative weighted networks.}
\label{fig:negative_cors_final}
\end{figure}

\end{document}